\documentstyle[12pt,axodraw]{article}

\begin{document}

\def\bra{\langle}
\def\ket{\rangle}
\def\beq{\begin{eqnarray}}
\def\eeq{\end{eqnarray}}
\def\qq{\langle \bar q q\rangle}
\def\qu{\langle \bar u u\rangle}
\def\qd{\langle \bar d d\rangle}
\def\qs{\langle \bar s s\rangle}
\def\qc{\langle \bar c c\rangle}
\def\qb{\langle \bar b b\rangle}

%{\bf On Mass Formulas for Charm and Beauty Baryons}
%{Массовые формулы для шармовых и бьюти- барионов}
\vskip 5mm
\title{On Mass Formulas for Charm and Beauty Baryons}
\author{T. M. Aliev$^a$
\thanks{Permanent institute: Institute of Physics, Baku, Azerbaijan}
\thanks{taliev@metu.edu.tr}
,~A. Ozpineci$^a$ \thanks{ozpineci@metu.edu.tr}
,~V. Zamiralov$^b$ \thanks{zamir@depni.sinp.msu.ru}
\\
{\small
$^a$Middle East Technical University, Ankara, Turkey} \\
{\small
$^b$Institute of Nuclear Physics, M. V. Lomonosov MSU, Moscow, Russia
}}
\begin{titlepage}
\maketitle
\thispagestyle{empty}

\begin{abstract}
\qquad Possible mixing and its consequences
of heavy cascade baryons $\Xi-\Xi^\prime$ %with the new quantum numbers
is analyzed and its importance in the analysis of their characteristics is shown
within the non-relativistic quark model and QCD sum rules.
%The quark model of \cite{Ono} is used as an example.
%Masses of new baryons as well as mixing angles of the states
%$\Xi-\Xi^\prime$ are obtained.
%The approach is applied to the interpolating currents
%of these baryons in the framework of the QCD sum rules.
\end{abstract}
\end{titlepage}
\section{Introduction}
During the last few years, very intriguing observations appeared in charm and
beauty baryon spectroscopy\cite{PDG08}. It would be interesting to discuss how successfully various models
describe the mass of the observed baryons. The central problem of all studies on this subject is
establishing the structure of new baryons within the quark model, in particular, of the cascade baryons with new flavors (see \cite{Ono}-\cite{RWol} and references therein).

On the other hand, many characteristics of the baryons are successfully determined in the framework of QCD sum rules method \cite{Ioffe}(see \cite{khodjamirian} about the modern status of this method).
The main issue in the applications of sum rules is the choice of an interpolating current with the same quantum numbers of the corresponding baryon.

In the present work, we discuss some interesting points changing considerably the usual way of constructing mass formulae for the heavy baryons either within the quark model or in
the QCD sum rules method.

The plan of this work is as follows. In section II, we discuss the problem of mixing of the baryon wave functions in the quark model \cite{Ono} and the obtained results
are used to construct wave functions of the baryons which are used to calculate the mass as well as the magnetic moments of the baryons. In the last section, we apply the same method
to construct interpolating QCD currents and modify the sum rules. In conclusion, we discuss our results briefly.
\section{ Mixing of the cascade baryons
$\Xi_c$- $\Xi_c^\prime$ in the quark model}

\qquad Let us consider standard non-relativistic quark model (NRQM) wave functions of the
charm cascade baryons
$\Xi_c$'s of the $SU(4)$ $20^\prime$-plet, having
in their content three different quarks $u$, $s$, $c$.
Upon reduction to
$SU(3)$ these baryons occur in the sum of sextet and triplet representations 
of $SU(3)$ ($20^\prime_4=8_3+6_3+\bar3_3+3_3$).
One can choose the wave functions of the $\Lambda-$ like
baryons with the quark
content $(usc)$ as
%20`-плет барионов о baryons with the usual
%quark wave functions
%\begin{equation}
%\sqrt{6}B^{\alpha}_{[\beta\gamma]}=\epsilon_{\beta\gamma\delta\rho}
%\{q^\alpha_1,q^\delta_1\}q^\rho,
%\end{equation}
%wherefrom we can define
\beq
%\sqrt{6}B^4_{[24]}
\sqrt{6}\Xi_c([us]c)=-c_1 u_1 s_2-u_1 c_1 s_2+s_1 c_1 u_2+c_1s_1u_2
\\
%\sqrt{6}B^3_{[23]}
\sqrt{6}\Xi_c([uc]s)=+s_1 u_1 c_2+u_1 s_1 c_2-s_1 c_1 u_2-c_1s_1u_2
\\
%\sqrt{6}B^1_{[21]}
\sqrt{6}\hat\Xi_c([sc]u)=-s_1 u_1 c_2-u_1 s_1 c_2+s_1 u_1 s_2+u_1c_1s_2
\label{lam1}
\eeq
Since the sum of these states is equal to zero,  only two of the states
are linearly independent. But any two of them are not orthogonal to
each other, so we construct to every state of the Eq.(\ref{lam1})
three possible orthogonal combinations,  corresponding
to $\Sigma$ like baryons:
\beq
%-\sqrt{6}(B^1_{[21]}-B^3_{[23]})
2\sqrt{3}\Xi^\prime_c(\{us\}c)=2s_1 u_1 c_2+2u_1 s_1 c_2
-c_1 u_1 s_2-u_1 c_1 s_2-s_1 c_1 u_2-c_1s_1u_2
\\
%-\sqrt{3}(B^4_{[24]}-B^1_{[21]})
2\sqrt{3}\Xi^\prime_c(\{uc\}s)=2c_1 u_1 s_2+2u_1 c_1 s_2
-s_1 u_1 c_2-u_1 s_1 c_2-s_1 c_1 u_2-c_1s_1u_2
%\quad (\Sigma-{\rm like})
\\
%\sqrt{6}(B^4_{[24]}-B^3_{[23]})
2\sqrt{3}\Xi^\prime_c(\{sc\}u)=2c_1 s_1 u_2+2s_1c_1u_2
-s_1 u_1 c_2-u_1 s_1 c_2-s_1 u_1 s_2-u_1c_1s_2
\label{sig2}
\eeq
In principle,
one can choose Eqs.(1,4), or (2,5) or (3,6)
as a pair of charm cascade baryons $\Xi_c,\Xi_c^{\prime}$. 
Rotating the Eqs.(1,4) by
$60^\circ$ one obtains (up to a sign)
the states (2,5); while rotating them
by $120^\circ$ one obtains the states (3,6) (up to a sign).

Upon reduction of the
$SU(4)$ $20^\prime$- pet to the SU(3) multiplets along the values of
charm, $C=0,1,2$, one can see that 
$\Lambda([ud]c^+)$, $\Xi_c^+([us]c)$ and $\Xi^0_c([ds]c)$ form
an anti-triplet while
$\Sigma(\{ud\}c)$, $\Xi_c^{\prime +}(\{us\}c)$ and 
$\Xi_c^{\prime 0}(\{ds\}c)$ enter the $SU(3)$ sextet.

However there is no reason to expect that the experimentally
observed charm cascade baryons $\Xi_c,\Xi_c^{\prime}$ should
belong to the pure anti-triplet or pure sextet states.
Upon choosing another pair of the states
$\Xi_c,\Xi_c^{\prime}$ from the Eqs.(1-6) these anti-triplet and
sextet states  mix similar to mixing of the
initial pure states of the
octet $\omega_0$ state and unitary singlet $\phi_0$ state
yielding the observed vector nonet mesons $\omega$ and $\phi$.

As an example, let us choose a pair of charm cascade baryons
$\Xi_c,\Xi_c^{\prime}$ with the quark content $(usc)$.
Formally, masses of the
$\Xi_c$ are defined as
\beq
M_{\Xi_c}=\bra \Xi_c| \hat m |\Xi_c\ket,
\quad M_{\Xi_c^\prime}=\bra \Xi_c^\prime| \hat m |\Xi_c^\prime\ket
\label{massme}
\eeq
where $\hat m$ is a mass operator in NRQM \cite{Ono}.

Using \cite{Altug} and the pair of states given in Eqs.(1) and (4),
masses of 
all $(usc)$ cascade baryons are related to each other
as
\beq
M_{\Xi^\prime_c(\{us\}c)}+3 M_{\Xi_c([us]c)}=
2 M_{\Xi^\prime_c(\{uc\}s)}+2 M_{\Xi^\prime_c(\{cs\}u)},
\nonumber\\
M_{\Xi_c([us]c)}+3 M_{\Xi^\prime_c(\{us\}c)}=
2 M_{\Xi_c([uc]s)}+ 2 M_{\Xi_c([cs]u)}.
\eeq
Analogous relations exist for other choices of the  cascade pair:
\beq
M_{\Xi^\prime_c(\{uc\}s)}+3 M_{\Xi_c([uc]s)}=
2 M_{\Xi^\prime_c(\{us\}c)}+2 M_{\Xi^\prime_c(\{cs\}u)},
\nonumber\\
M_{\Xi_c([uc]s)}+3 M_{\Xi^\prime_c(\{uc\}s)}=
2 M_{\Xi_c([us]c)}+ 2 M_{\Xi_c([cs]u)}
\eeq
and 
\beq
M_{\Xi^\prime_c(\{cs\}u)}+3 M_{\Xi_c([cs]u)}=
2 M_{\Xi^\prime_c(\{uc\}s)}+2 M_{\Xi^\prime_c(\{us\}c)},
\nonumber\\
M_{\Xi_c([cs]u)}+3 M_{\Xi^\prime_c(\{cs\}u)}=
2 M_{\Xi_c([uc]s)}+ 2 M_{\Xi_c([us]c)}.
\eeq

In general, the off-diagonal terms are not zero.
Upon using relations from \cite{Altug}
non-diagonal mass terms of the pair
$\Xi_c([us]c)$ и $\Xi^\prime_c(\{us\}c)$ could be written in terms
of other states of the Eqs.(1-6) as
$$
\sqrt3 M_{\Xi_c([us]c)\Xi^\prime_c(\{us\}c)}=
M_{\Xi^\prime_c(\{uc\}s)}- M_{\Xi^\prime_c(\{cs\}u)}=
-M_{\Xi_c([uc]s)}+ M_{\Xi_c([cs]u)},
$$
and similarly for the other two pairs of the heavy cascade baryons:
$$
\sqrt3 M_{\Xi_c([uc]s)\Xi^\prime_c(\{uc\}s)}=
-M_{\Xi^\prime_c(\{us\}c)}+ M_{\Xi^\prime_c(\{cs\}u)}=
M_{\Xi_c([us]c)}-M_{\Xi_c([cs]us)},
$$
$$
\sqrt3 M_{\Xi_c([cs]u)\Xi^\prime_c(\{cs\}u)}=
M_{\Xi^\prime_c(\{us\}c)}- M_{\Xi^\prime_c(\{uc\}s)}=
- M_{\Xi_c([us]c)}+M_{\Xi_c([uc]s)}.
$$

%These non-diagonal 
%mass terms in general could be not zero, that is, 
%$$
%M_{\Xi_c \Xi_c^\prime}=\bra \Xi_c| \hat m |\Xi_c^\prime\ket\not =0.
%$$
It is rather obvious that only for some particular choice of
parameters the non-diagonal mass terms are equal to zero. For example, isotopic invariance leads to the vanishing of these
non-diagonal mass terms of the baryons
$\Lambda([ud]h^)$ and $\Sigma(\{ud\}h)$, $h=s,c,b$.

It is clear that the states of definite mass and their masses can be obtained by diagonalizing the mass matrix:
%But what masses should we choose then to compare them with experiment?
%First of all they should be diagonal in a given model.
%To answer this question in some detail let us write a mass 
%2$\times$2 matrix 
of the chosen model \cite{Ono}
\beq
\hat M_{\Xi_c}=\left(\begin{array}{cc}
M_{\Xi_c^\prime} & M_{\Xi_c\Xi_c^\prime}
\\M_{\Xi_c^\prime\Xi_c} & M_{\Xi_c}
\end{array}\right).\quad
\eeq
The corresponding  secular equation yields the physical masses to be:
\beq
M_{\Xi_c}^{1,2}=\frac{1}{2}(M_{\Xi_c}+M_{\Xi_c^\prime})\pm
\frac{1}{2}\sqrt{(M_{\Xi_c}-M_{\Xi_c^\prime})^2+
4 M_{ \Xi_c\Xi_c^\prime}^2},
\label{masseigenvalues}
\eeq
where the off-diagonal elements are assumed to be equal, i.e. $ M_{ \Xi_c\Xi_c^\prime}$ =$M_{\Xi_c^\prime\Xi_c}$.
(Recently this formula was written also in
\cite{Karliner})
It is these masses which one should compare with experiment.
The values of these masses do not depend on which pair from Eqs. (1-6) is chosen for the $\Xi_c$ and $\Xi_c^\prime$ states
%upon the mode of
%grouping quarks in "diquarks", 
since the sum of
$M_{\Xi_c}$ and $M_{\Xi_c^\prime}$ as well as the
square root in Eq. \ref{masseigenvalues}  are invariant under rotations in the
flavor space by $60^\circ$ and $120^\circ$.

Thus for a chosen representation of the $\Xi_c$ and $\Xi_c^\prime$ baryons,
if the off-diagonal entries are not zero, i.e.
%if in some model mass non-diagonal term is different from zero,
$ M_{ \Xi_c\Xi_c^\prime}$=$\bra \Xi_c|\hat m|\Xi_c^\prime \ket\not=0$,
then the corresponding wave functions do not describe
observable particles. To obtain the representation of the observable particles,
the $\Xi_c$ and $\Xi_c^\prime$ should be rotated by some angle
$\alpha$,
\begin{eqnarray}
\Xi_c^{\prime\alpha}= \Xi_c^\prime \cos_\alpha+\Xi_c \sin \alpha,\qquad
%\\
\Xi_c^\alpha= -\Xi_c^\prime \sin\alpha +\Xi_c \cos\alpha.
\end{eqnarray}
Requiring that the off-diagonal elements of the mass matrix for these newly defined states to be zero, 
%Perfoming calculations of the matrix elements of the
%mass operator between the rotated states 
%after some algebra one obtains that non-diagonal mass term
%goes to zero at the angle $\alpha$ given by the relation
the rotation angle $\alpha$ should be chosen such that:
\begin{equation}
\tan 2\alpha = \frac{2 \bra \Xi_c^\prime |\hat m |\Xi_c\ket}
{(M_{\Xi_c^\prime}-M_{\Xi_c})},
\end{equation}
where $\Xi_c^\prime$, $\Xi_c$ are any pair (1,4), (2,5), or (3,6)
from the Eqs.(1-6).%f{sig2}) of the model.
It would be natural
to define these  diagonalized states as the physical ones.
In this case a quark structure of the baryon (in the given model!)
would be a superposition of the states
$\Xi_c$ and $\Xi_c^\prime$.

\section{Quark model for masses of new baryons}

\qquad Let us apply the approach presented in the previous section 
heavy baryons within the model of \cite{Ono}. It is convenient to use the fact that already 20
years ago masses of heavy baryons with new quantum numbers
were calculated within quark models
(cf., e.g., \cite{Ono}-\cite{Verma}).
It is of interest to note that the predictions on the masses
of heavy baryons of these models are in surprisingly
good agreement with the modern data.
In this work, the mass operator of \cite{Ono} is used due to its simplicity and
clearness:
\beq
M_B=m_0+\sum_{i=1}^3 m_i +\chi \sum _{i>j}
\frac{\vec S_i\cdot \vec S_j}{m_i\cdot m_j},
\label{ono}
\eeq
where $m_0$ is an overall constant, $m_0=77$ MeV, and 
$\chi=22.05\cdot 10^{-3}~ GeV^3$ \cite{Ono};
$S_q$ is the spin operator of the quark $q$.
Quark masses are taken from \cite{Ono}:
$$
m_u=m_d=336\, {\rm MeV}, \quad m_s=510\, {\rm MeV},
\quad m_c=1680\, {\rm MeV},\quad m_b=5000\, {\rm MeV}.
$$

%{\bf Charm baryons}

%\vskip 3mm
%{\bf Charm and beauty cascade hyperons}
%{\bf Шармовые и бьюти каскадные барионы}
\vskip 3mm
Masses of charm and beauty cascade and $\Omega^-$ baryons were
calculated using  Eq.(\ref{ono}) in \cite{Ono} for particular
quark combinations of baryons.
We have performed calculations for all the possible
quark combinations in baryons with the same formula and put
results into the Table 1.

%{\bf Double-heavy baryons}

As an example, the analysis  for baryons
with two different heavy quarks forming $SU(3)$ triplets
($\Xi_{cb}^{\prime +,0}$, $\Omega_{cb}^{\prime 0}$)
and ($\Xi_{cb}^{+,0}$, $\Omega_{cb}^{0}$) is presented below.
We would write in detail calculations for the pair 
$\Xi_{cb}^{+}$ and $\Xi_{cb}^{\prime +}$.
%{\bf CASCADE cb-BARYONS}

(1) We begin with the quark content proposed in \cite{Ono}, i.e.
let $b$ be a single quark while the pair $(uc)$
in (anti)symmetric state form a diquark.
Then, for the diagonal elements of the mass matrix, we get
$$
M_{\Xi_{cb}([cu]b)}=m_0+m_u+m_c+m_b+\chi(-\frac{3}{4})\frac{1}{m_u\cdot m_c}=
%$$
%$$
%=7093+22.05(-\frac{3}{4}) 2.976\cdot 2.796/5=
%7093-29.3=
7063.7\,{\rm МэВ};(7064 \cite{Ono})
$$
$$
M_{\Xi_{cb}^\prime(\{cu\}b)}=m_0+\sum_q m_q+\chi(-\frac{1}{2}\frac{1}{m_u m_b}
-\frac{1}{2}\frac{1}{m_c m_b}+\frac{1}{4}\frac{1}{m_u m_c})=
%$$
%$$
%=7093-6.5625-1.3125+9.7656=7093+1.891=
7094.9 (7095 \cite{Ono})
$$
%$$
%\frac{1}{2}(\Xi_{cb}(\{cu\}b)+\Xi_{cb}^\prime(\{cu\}b))=7079.3;
%$$
%$$
%(\Xi_{cb}(\{cu\}b)-\Xi_{cb}^\prime(\{cu\}b))=-31.2;
%$$
and the non diagonal matrix elements are:
$$
|\sqrt{3}\bra\Xi_{cb}^\prime(\{cu\}b)\Xi_{cb}([cu]b) \ket|=
\chi\frac{3}{4}(\frac{1}{m_u }-\frac{1}{m_c})\frac{1}{ m_b}
\sim 7.9\,{\rm MeV}
$$
%$$
%\Xi_{cb}^\prime(\{cu\}b)\quad{\rm >}\quad \Xi_{cb}([cu]b).
%$$
Resolving the secular equation
$$
x_{1,2}=7079.3\pm\frac{1}{2}\sqrt{31.2^2+\frac{4}{3}7.9^2}=
7079.3\pm 16.4,
$$
one obtains the mass eigenvalues as
$$
x_1=7095.7\qquad x_2=7063.5
$$
In order to go from the initial states 
$\Xi$, $\Xi^\prime$ to those with the masses
$x_{1,2}$ one should rotate them at the angle $\alpha=16.3^\circ/2=8.15^o$,
$\tan 2\alpha=0.2924$.

(2) Now, let the  quarks
$c$ and $b$ form a diquark while the light quark $u$
is the single one. In this case
$$
M_{\Xi_{cb}([cb]u)}=m_0+m_u+m_c+m_b+
\chi(-\frac{3}{4})\frac{1}{m_b\cdot m_c}=
%$$
%$$
%=7093+22.05(-\frac{3}{4}) 1/5\cdot 2.796/5=
7091\,{\rm MeV},
$$
$$
M_{\Xi_{cb}^\prime(\{cb\}u)}=m_0+\sum_q m_q+\chi(-\frac{1}{2}\frac{1}{m_u m_c}
-\frac{1}{2}\frac{1}{m_u m_b}+\frac{1}{4}\frac{1}{m_b m_c})=
%$$
%$$
%=7093-19.53-6.5625+0.65625=7093-25.43625=
7067.6\,{\rm MeV}.
$$
%$$
%\frac{1}{2}(\Xi_{cb}(\{cb\}u)+\Xi_{cb}^\prime(\{cb\}u))=7079.3;
%$$
%$$
%(\Xi_{cb}(\{cb\}u)-\Xi_{cb}^\prime(\{cb\}u))=23.4;
%$$
The non-diagonal mass matrix element is now
$$
|\sqrt{3}\bra\Xi_{cb}^\prime(\{cb\}u)\Xi_{cb}([cb]u) \ket|=
\chi\frac{3}{4}(\frac{1}{m_c }-\frac{1}{m_b})\frac{1}{ m_u}
\sim 19.4\,{\rm MeV},
$$
%$$
%\Xi_{cb}([cb]u)\quad{\rm >}\quad \Xi_{cb}^\prime(\{cb\}u)
%$$
wherefrom the eigenvalues are obtained as follows 
$$
x_1=7095.2\,{\rm MeV}\qquad x_2=7062.8\,{\rm MeV}.
$$
In order to go from the initial states 
$\Xi$, $\Xi^\prime$ to those with the masses
$x_{1,2}$ one should rotate them at the angle 
$\alpha=136.25^o/2=68.13^o$,
$\tan 2\alpha=-0.9573$.

(3) Now quarks $u$ and $b$ form a diquark while $c$ is single.
Then, the masses of the cascade baryons are given by:
$$
M_{\Xi_{cb}([ub]c)}=m_0+m_u+m_c+m_b+
\chi(-\frac{3}{4})\frac{1}{m_b\cdot m_u}=
%$$
%$$
%=7093+22.05(-\frac{3}{4}) 1/5\cdot 2.796=7093-9.9=
7083.1\,{\rm MeV};
$$
$$
M{\Xi_{cb}^\prime(\{ub\}c)}=m_0+\sum_q m_q+(-\frac{1}{2}\frac{1}{m_u m_c}
-\chi\frac{1}{2}\frac{1}{m_b m_c}+\frac{1}{4}\frac{1}{m_b m_u})=
$$
$$
%=7093-27.7=
7075.3\,{\rm MeV};
$$
%$$
%\frac{1}{2}(\Xi_{cb}(\{ub\}c)+\Xi_{cb}^\prime(\{ub\}c))=7079.2;
%$$
%$$
%(\Xi_{cb}(\{ub\}c)-\Xi_{cb}^\prime(\{ub\}c))=7.7;
%$$
The non-diagonal mass matrix element being
$$
|\sqrt{3}\bra\Xi_{cb}^\prime(\{cb\}u)\Xi_{cb}([cb]u) \ket|=
\chi\frac{3}{4}(\frac{1}{m_u }-\frac{1}{m_b})\frac{1}{ m_c}
\sim 27.7\,{\rm MeV}
$$
%$$
%\Xi_{cb}([cb]u)\quad{\rm >}\quad \Xi_{cb}^\prime(\{cb\}u)
%$$
with the solutions to the secular equation
%$$
%x_{1,2}=7079.3\pm\frac{1}{2}\sqrt{7.7^2+\frac{4}{3}27.7^2}=
%7079.3\pm 16.4\,{\rm MeV}
%$$
%and solutions
$$
x_1=7095.7\,{\rm MeV}\qquad x_2=7062.9\,{\rm MeV}.
$$
In order to go from the initial states 
$\Xi$, $\Xi^\prime$ to those with the masses
$x_{1,2}$ one should rotate them at the angle 
$\alpha=256.5^o/2=128.25^o$,
$\tan 2\alpha=4.1539$.

%We note also here that rotating by 60$^o$ we arrive from (1) to (2) and then
%to (3) as it has been written above.

Similar calculations are done
also for other charm and beauty baryons and the results are
presented in Table 1.
 
From our analysis, it
follows that the predictions on the mass of the physically
observed baryons does not depend on the
particular quark construction of the baryons in the given model.

%However the wave function of the baryon with the diagonal mass matrix
%as well as all the characteristics of the baryon such as
%magnetic moments, weak decay constants, strong couplings and etc.
%depend upon these particular construction through the angle
%$\alpha$ mixing the initial states $\Xi$ and $\Xi^\prime$.

\subsection*{Magnetic moments of the double-flavored baryons}

\qquad As another application of our approach, let us consider the 
magnetic moments of the double-flavored baryons
$\Xi_{cb}^{\prime +}$ and $\Xi_{cb}^{+}$.

Magnetic moment of the baryon with the quark content
$\Xi_{cb}^{+}([cb]u)$, which is used often in the modern works
(cf., e.g., \cite{Faust}) in the NRQM, is equal to
the magneton of just the $u$ quark, $\mu_u$, while that of the baryon
$\Xi_{cb}^{\prime +}(\{cb\}u)$ is  equal to
$(2\mu_c+2\mu_b-\mu_u)/3$.

However baryons of such quark content in the model of
\cite{Ono} have large non-diagonal mass terms, so one should
choose as the wave function of the baryon a linear combination
of $\Xi_{cb}^{\prime +}(\{cb\}u)$ and $\Xi_{cb}^{+}(\{cb\}u)$
with the mixing angle
$\alpha=68.13^\circ$ (cf. Table 1).
Then for the magnetic moments we have:
\begin{eqnarray} 
\bra\Xi_{cb}^{+\alpha}|\hat \mu|\Xi_{cb}^{+\alpha}\ket &=&
\frac{1}{3}(2\mu_c+2\mu_b-\mu_u) \sin^2\alpha+
\mu_u \cos^2 \alpha
- \frac{2}{\sqrt3} \sin\alpha \cos\alpha (\mu_c-\mu_b)
\nonumber \\
&=&
\left(1-\frac{4}{3} \sin^2\alpha \right)\mu_u=-0.148 \mu_u
\\
\bra\Xi_{cb}^{\prime +\alpha}|\hat \mu|\Xi_{cb}^{+\alpha}\ket&=&
\frac13 (2\mu_c+2\mu_b-\mu_u) \cos^2\alpha+
\mu_u \sin^2 \alpha
+ \frac{2}{\sqrt3} \sin\alpha \cos\alpha (\mu_c-\mu_b)
\nonumber \\
&=& \left(-\frac13+\frac43 \sin^2\alpha\right)\mu_u=0.815 \mu_u
\end{eqnarray} 
where we have neglected $\mu_b$ and $\mu_c$ in comparison to $\mu_u$.
%Отметим, что из теоретико-групповых соображений следует, что
These predictions do not depend on the particular choice of the
baryon pair $\Xi_{cb}-\Xi^\prime_{cb}$ provided one takes
the mixing angle $\alpha$ corresponding to the
given combination, and differ considerably from the
predictions given by the "pure" states
$\Xi_{cb}^{+}(\{cb\}u)$ and $\Xi_{cb}^{\prime +}(\{cb\}u)$
(cf. Table 2).

\section{ Mixing of the states
$\Xi-\Xi^\prime$ in QCD sum rules}

\qquad The problem of the $\Xi-\Xi^\prime$ mixing can also be analyzed within the QCD sum rules framework. The main difference in this case
is that, rather than working with the mass matrix, one deals with correlation functions of interpolating currents, i.e.
%in this case it relates more with the
%interpolating currents. 
%The final form of the interpolating
%current is dictated by the possible $\Xi-\Xi^\prime$ mixing.
%It means that prior to performing calculations of the
%magnetic moments, meson couplings etc. one should be assured that
%the corresponding interpolating currents yield the diagonal
%mass matrix for the $\Xi$ and $\Xi^\prime$ states.
 the role of matrix elements (\ref{massme})
is played by correlators \cite{Ioffe}
\beq
\Pi^{\Xi, \Xi^\prime}=
i \int d^4 x e^{ipx}\bra 0|T\{\eta(x)^{\Xi, \Xi^\prime}
\eta(0)^{\Xi, \Xi^\prime}\}|0\ket,
\eeq
which are calculated first in QCD  using OPE,
and secondly by inserting a complete set of physical states. Performing Borel transformation to
suppress high--excited states and equating both expansions one
obtains QCD sum rules \cite{Ioffe}.

In this case, mixing causes the non-diagonal correlation functions to have non-zero values. Prior to performing calculations of the
physical properties, such as the magnetic moments, meson couplings etc., of the baryons, one should 
make sure that the corresponding interpolating currents have zero non-diagonal correlators.

To find the combination of $\eta_\Xi$ and $\eta_{\Xi'}$ that have zero non-diagonal correlator, consider the following interpolating currents that
are obtained from $\eta_\Xi$ and $\eta_{\Xi'}$ after a rotation by $\alpha$:
\beq
\eta^\prime_\alpha=\eta_{\Xi^\prime}\cos\alpha+\eta_\Xi\sin\alpha,
\eta_\alpha=-\eta_{\Xi^\prime}\sin\alpha+\eta_\Xi\cos\alpha,
\eeq
where the mixing angle $\alpha$ should be chosen such that:
%should be done in a way that the OPE part of the QCD sum rules
%based on the non-diagonal correlator of the form
\beq
\Pi^{\Xi \Xi^\prime}_\alpha=
i \int d^4 x e^{ipx}\bra 0|T\{\eta_\alpha(x)
\eta^\prime_\alpha(0)\}|0\ket = 0,
\eeq
goes to zero.

Let us study this problem on a simplified toy-model
using the QCD mass sum rules \cite{Zu} written for the octet hyperons
$\Sigma^{0}$ and $\Lambda$. Generalization to our case seems not to be
without problems but is sufficient for our purposes.

Omitting the vacuum expectation values
of the quarks $c$ and $b$ and neglecting mass of the $u$ quark in the QCD sum rules, 
we construct following  \cite{Zu} the QCD mass sum rule for
the baryon $\Xi_{cb}^\prime(\{ub\}c)$
% a toy-model.
%===============================================================
\begin{eqnarray}
\frac{M^{6}}{8L^{4/9}}E_{2}+\frac{bM^{2}}{32L^{4/9}}E_{0}+
\frac{M^2}{4L^{4/9}}a_{u}m_{b}E_{0}-
\frac{m_0^2}{48L^{26/27}}3a_u m_{b}=
\beta^{2}_{\Xi^\prime}e^{-(M^{2}_{\Xi^\prime}/M^{2})},
\label{xiprim}
\end{eqnarray}
where $a_{q}$, $b$ and $a_{q}m_{0}^{2}$ are defined as
\cite{Ioffe}:
\begin{eqnarray}
a_{q}=-(2\pi)^{2}\qq,\quad
b=\langle g_{c}G^{2}\rangle,\quad 
L=ln(M^{2}/\Lambda^{2})/ln(\mu^{2}/\Lambda^{2})
\nonumber\\
a_{q}m_{0}^{2}=(2\pi)^{2}\langle g_{c}\bar q \sigma\cdot G q
\rangle,\quad q=u,d,s,
\label{vev}
\end{eqnarray}
$\mu$ being renormalization point, while $G$ is a gluon field
with the coupling $g_c$ to quarks.

The corresponding sum rule for the
$\Lambda$- like baryon $\Xi_{cb}([ub]c)$ reads:
%(upon neglecting vev of $c$ and $b$ quarks and mass of the
%$u$ quark)
\begin{eqnarray}
\frac{M^{6}}{8L^{4/9}}E_{2}+\frac{bM^{2}}{32L^{4/9}}E_{0}+
\frac{M^{2}}{12L^{4/9}}E_{0}a_u(2m_{c}-m_b)
\\
-\frac{m_0^2}{48L^{26/27}}a_u(2m_c-4m_b)
=\beta^{2}_{\Xi}e^{-(M^{2}_{\Xi}/M^{2})}.
\nonumber
\label{xi}
\end{eqnarray}
%The left-handed side (LHS) of these sum rules is equal to
%\beq
%(\frac{M^{2}}{6L^{4/9}}E_{0}
%-\frac{m_0^2}{24L^{26/27}})a_u(2m_b-m_c)
%%=\beta^{2}_{\Xi}e^{-(M^{2}_{\Xi}/M^{2})}.
%\nonumber
%\label{xidif}
%\eeq
%The LHS of the QCD sum rule corresponding to the
%mass non-diagonal transition between $\Xi$ and $\Xi^\prime$ reads:
The non-diagonal correlation function can also be written as:
\beq
\sqrt{3} \Pi^{\Xi\Xi^\prime}=
(\frac{M^{2}}{6L^{4/9}}E_{0}
-\frac{m_0^2}{24L^{26/27}})\frac{3}{2}a_u m_c.
\eeq
Requiring that $\Pi^{\Xi\Xi^\prime}_\alpha = 0$ leads to the value of the mixing angle given by:
$$
\tan 2\alpha _{(ub)c}=\frac{\sqrt{3}m_c}{2m_b-m_c}\sim 0.342 .
$$
For the other pair of heavy baryons
$\Xi_{cb}^\prime(\{uc\}b)$ and
$\Xi_{cb}([uc]b)$ one should just change $c\rightarrow b$, 
wherefrom
$$
\tan 2\alpha _{(uc)b}=\frac{\sqrt{3}m_b}{m_b-2m_c}\sim 5.18.
$$
Finally for $\Xi_{cb}^\prime(\{cb\}u)$ one obtains
\begin{eqnarray}
\frac{M^{6}}{8L^{4/9}}E_{2}+\frac{bM^{2}}{32L^{4/9}}E_{0}+
\beta^{2}_{\Xi^\prime}e^{-(M^{2}_{\Xi^\prime}/M^{2})},
\label{xiprim2}
\end{eqnarray}
while for $\Xi_{cb}([cb]u)$
\begin{eqnarray}
\frac{M^{6}}{8L^{4/9}}E_{2}+\frac{bM^{2}}{32L^{4/9}}E_{0}+
+\frac{2M^{2}}{12L^{4/9}}E_{0}a_u(m_{c}+m_b)
\\
-\frac{2m_0^2}{48L^{26/27}}a_u(m_c+m_b)
=\beta^{2}_{\Xi}e^{-(M^{2}_{\Xi}/M^{2})},
\nonumber
\label{xi2}
\end{eqnarray}
wherefrom
$$
\tan 2\alpha _{(cb)u}=\frac{\sqrt{3}(m_c-m_b)}{(m_c+m_b)}\sim -0.872
$$
These formulae are transformed from one into the other
by shifting the angle $\alpha$ by $60^\circ$ and $120^\circ$.
At $m_c$=1650 MeV, $m_b$=5 GeV one obtains
$\alpha_{(ub)c}\sim 9.5^o$,
$\alpha_{(cb)u}\sim 69.5^o$, $\alpha_{(uc)b}\sim 129.5^o$.

It is of interest to note that mass relations of QCD in this approximation
yield somewhat unexpected result that the minimal mixing angle
favors the diquark pair 
$(ub)$ while others lead to large mixing angles.

Calculation of the magnetic moments of the double-flavored baryons
in the quark model with the mixing angles from the  QCD toy-model
yield practically the same results as the quark model of \cite{Ono}.

\section{Conclusion}

\qquad We have tried to show the importance of mixing
of heavy cascade baryons $\Xi-\Xi^\prime$ %with the new quantum numbers
in analysis of their characteristics.
%It is only by chance that wave functions of heavy baryons $\Xi$
%could lead to the diagonal mass matrix.
As an example, the non-relativistic quark model of \cite{Ono} is used.
The same approach is applied to the interpolating currents
of these baryons in the framework of the QCD sum rules
which is shown on the example of simplified mass QCD sum rules.
The main conclusion is that in any given model of heavy baryons,
one should first find the quark configuration leading to vanishing diagonal
matrix elements. After finding the physical states, 
one should perform calculations of
the various characteristics of these baryons.
In the case of the QCD sum rules where there is no mass formulae
in the common sense of the word, the problem of the truthful
combination of interpolating currents deserves
further study.

\section*{Acknowledgement}
T.M.A. and A.O. acknowledge the support of the European Community-Research Infrastructure
Integrating Activity
┐Study of Strongly Interacting Matter┐ (acronym HadronPhysics2, Grant Agreement
n. 227431)
under the Seventh Framework Programme of EU.
\newpage
Table 1. Baryon masses (in MeV) and mixing angles in the Ono model
\vskip 5mm
\begin{tabular}{|c|c|c|c|c|c|c|} \hline
&&&&&& \\ 
Baryon & $\{q_1 q_2\}q_3$  & $[q_1 q_2]q_3$
& $\sqrt3\bra\Xi^\prime|\hat m|\Xi\ket$
& $x_1$ & $x_2$ & $\alpha$ 
\\  
&&&&&&
\\ \hline
&&&&&& \\ 
$ \Xi_{c}((us)c) $    & 2603  & 2504 & 9.9 & 2603  & 2503 & 3.3$^o$
\\ 
&&&&&&\\
\hline 
&&&&&& \\ 
$ \Xi_{c}((sc)u) $    & 2523.8 & 2558 & 69.3 & 2603
& 2503 & 63.25$^o$
\\
&&&&&&\\ 
\hline
&&&&&& \\ 
$ \Xi_{c}((uc)s) $    & 2534  & 2573.3 & 79.2 & 2603
& 2503 & 123.3$^o$
\\
&&&&&&\\
\hline
&&&&&& \\ 
$ \Xi_{b}((us)b) $    & 5945  & 5824 & 3.3 & 5945
& 5824 & 0.9$^o$
\\
&&&&&& \\
\hline
&&&&&& \\ 
$ \Xi_{b}((sb)u) $    & 5852.6  & 5916.4 & 89.1 & 5944
& 5824 & 60.8$^o$
\\ 
&&&&&& \\ 
\hline 
&&&&&& \\ 
$ \Xi_{b}((ub)s) $    & 5855.9  & 5913.1 & 92.4 & 5945
& 5824 & 120.9$^o$
\\ 
&&&&&& \\ 
\hline
&&&&&& \\ 
$ \Xi_{cb}((uc)b) $    & 7094.9  & 7963.7 & 7.9 & 7095.7
& 7063.5 & 8.15$^o$
\\ 
&&&&&& \\ 
\hline 
&&&&&& \\ 
$ \Xi_{cb}((cb)u) $    & 7067.6  & 7091.0 & 19.4 & 7095.2
& 7062.8 & 68.13$^o$
\\ 
&&&&&& \\ 
\hline
&&&&&& \\ 
$ \Xi_{b}((ub)s) $    & 7075.3  & 7083.1 & 27.7 & 7095.9
& 7062.9 & 128.25$^o$
\\ 
&&&&&& \\ 
\hline
&&&&&& \\ 
$ \Omega_{cb}((sc)b) $    & 7267.9  & 7247.2 & 4.62 & 7268.24
& 7246.86 & 7.22$^o$
\\ 
&&&&&& \\ 
\hline 
&&&&&& \\ 
$ \Omega_{cb}((cb)s) $    & 7250.0  & 7265.0 & 13.2 & 7268.24
& 7246.86 & 67.27$^o$
\\ 
&&&&&& \\  
\hline 
&&&&&& \\ 
$ \Omega_{cb}((sb)c) $    & 7254.7  & 7260.4 & 17.82 & 7268.23
& 7246.87 & 127.26$^o$
\\ 
&&&&&& \\
\hline
\end{tabular}

\newpage

Table 2. Magnetic moments of double-flavored baryons
in NRQM for "pure" and mixed states (with index $\alpha$)
\vskip 5mm
\begin{tabular}{|c|c|c|} \hline
&& \\ 
Baryon & NRQM  &  NRQM$\alpha$
\\  
&&
\\ \hline
&& \\ 
$ \Xi_{cb}([cb]u) $    & $\mu_u$  & -0.148$\mu_u$ 
\\ 
&&\\
\hline 
&& \\ 
$ \Xi_{cb}(\{cb\}u) $    & -1/3$\mu_u$  & 0.815$\mu_u$ 
\\ 
&&\\
\hline
&& \\ 
$ \Xi_{cb}([cu]b) $    & $\mu_b\sim 0$  & -0.148$\mu_u$ 
\\ 
&&\\
\hline 
&& \\ 
$ \Xi_{cb}(\{cu\}b) $    & 2/3$\mu_u$  & 0.815$\mu_u$ 
\\ 
&&\\
\hline 
&& \\ 
$ \Xi_{cb}([ub]c) $    & $\mu_c\sim 0$  & -0.148$\mu_u$ 
\\ 
&&\\
\hline 
&& \\ 
$ \Xi_{cb}(\{ub\}c) $    & 2/3$\mu_u$  & 0.815$\mu_u$ 
\\ 
&&\\
\hline 
\end{tabular}

\end{document}